\documentclass[aps,prb,preprint,superscriptaddress]{revtex4-2}

\usepackage{gensymb}
\usepackage{graphicx}
\usepackage{amsmath}
\usepackage{amssymb}
\usepackage{epstopdf}

\begin{document}


	
\title{Optical magnons with dominant bond-directional exchange interactions in a honeycomb lattice iridate $\alpha$-Li$_{2}$IrO$_{3}$}



\author{Sae Hwan Chun}
\email{pokchun81@gmail.com}
\affiliation{Materials Science Division, Argonne National Laboratory, Argonne, Illinois 60439, USA}
\affiliation{Department of Physics, University of Toronto, 60 St. George Street, Toronto, ON M5S 1A7, Canada}
\affiliation{Pohang Accelerator Laboratory, Pohang, Gyeongbuk 37673, Republic of Korea}
\author{P. Peter Stavropoulos}
\affiliation{Department of Physics, University of Toronto, 60 St. George Street, Toronto, ON M5S 1A7, Canada}
\author{Hae-Young Kee}
\affiliation{Department of Physics, University of Toronto, 60 St. George Street, Toronto, ON M5S 1A7, Canada}
\affiliation{Canadian Institute for Advanced Research, Toronto, Ontario, M5G 1Z8, Canada}
\author{M. Moretti Sala}
\affiliation{ESRF, The European Synchrotron, 71 Avenue des Martyrs, F-38000 Grenoble, France}
\affiliation{Dipartimento di Fisica, Politecnico di Milano, Piazza Leonardo da Vinci 32, I-20133 Milano, Italy}
\author{Jungho Kim}
\affiliation{Advanced Photon Source, Argonne National Laboratory, Argonne, Illinois 60439, USA}
\author{Jong-Woo Kim}
\affiliation{Advanced Photon Source, Argonne National Laboratory, Argonne, Illinois 60439, USA}
\author{B. J. Kim}
\affiliation{Department of Physics, Pohang University of Science and Technology, Pohang, Gyeongbuk 37673, Republic of Korea}
\affiliation{Center for Artificial Low Dimensional Electronic Systems, Institute for Basic Science (IBS), 77 Cheongam-Ro, Pohang, Gyeongbuk 37673, Republic of Korea}
\author{J. F. Mitchell}
\affiliation{Materials Science Division, Argonne National Laboratory, Argonne, Illinois 60439, USA}
\author{Young-June Kim}
\affiliation{Department of Physics, University of Toronto, 60 St. George Street, Toronto, ON M5S 1A7, Canada}




\begin{abstract}
We have used resonant inelastic x-ray scattering to reveal optical magnons in a honeycomb lattice iridate $\alpha$-Li$_{2}$IrO$_{3}$. The spectrum in the energy region 20-25 meV exhibits momentum dependence, of which energy is highest at the location of the magnetic Bragg peak, (\textit{h}, \textit{k}) = ($\pm$0.32, 0), and lowered toward (0, 0) and ($\pm$1, 0). We compare our data with a linear spin-wave theory based on a generic nearest-neighbor spin model. We find that a dominant bond-directional Kitaev interaction of order 20 meV is required to explain the energy scale observed in our study. The observed excitations are understood as stemming from optical magnon modes whose intensity is modulated by a structure factor, resulting in the apparent momentum dependence. We also observed diffuse magnetic scattering arising from the short-range magnetic correlation well above $\textit{T}_{N}$. In contrast to Na$_{2}$IrO$_{3}$, this diffuse scattering lacks the $C_3$ rotational symmetry of the honeycomb lattice, suggesting that the bond anisotropy is far from negligible in $\alpha$-Li$_{2}$IrO$_{3}$.
\end{abstract}



\maketitle

\clearpage
The Kitaev quantum spin liquid (QSL) state emerges out of the magnetic frustration driven by bond-directional Ising-type interactions between \textit{S} = 1/2 spins on a honeycomb lattice (referred to as Kitaev interactions) \cite{Kitaev, Rau01, Hermanns, Takagi}. This quantum state has gained intense interest with the prospect that it can host novel magnetic excitations such as Majorana fermions -- non-Abelian anyons with promising application potentials for quantum computing \cite{Kitaev}. The search for a realization of the Kitaev QSL began with honeycomb iridates $\textit{A}$$_{2}$IrO$_{3}$ ($\textit{A}$ = Na, Li) \cite{Jackeli, Chaloupka01, Singh01, Singh02}. The combination of strong spin-obit coupling, electronic correlation, and cubic crystal field in Ir$^{4+}$ (5$d^{5}$) leads to a $\textit{J}_{eff}$ = 1/2 state that has a pseudospin 1/2 moment \cite{BJKim01, BJKim02}. The edge-shared IrO$_{6}$ octahedra contribute to producing the bond-directional Kitaev interaction while suppressing the otherwise dominant Heisenberg interaction (Fig. 1(a)) \cite{Jackeli}. However, the ground state of the honeycomb iridates Na$_{2}$IrO$_{3}$ and $\alpha$-Li$_{2}$IrO$_{3}$ as well as their 3D extension to ‘hyper-honeycomb ($\beta$-)’ and ‘harmonic-honeycomb ($\gamma$-)’ Li$_{2}$IrO$_{3}$ has long-range magnetic order as a consequence of isotropic Heisenberg and other symmetry-allowed exchange interactions \cite{Singh01, Singh02, Liu, Ye, Modic, Takayama}. It turns out that excited states still can bear the Kitaev QSL physics \cite{Chun, Sandilands, Glamzada, Nasu, Banerjee, SHDo, Gohlke, Kitagawa, Kasahara} as exemplified in Na$_{2}$IrO$_{3}$ \cite{Chun}, $\alpha$-RuCl$_{3}$ \cite{Banerjee, SHDo, Kasahara}, and H$_{3}$LiIr$_{2}$O$_{6}$ \cite{Kitagawa}, but its relevance is challenged by the influence of the other exchange interactions.

Quantitative investigation of the exchange interactions in known honeycomb magnets has not been carried out successfully to date. One of the difficulties is that interpretation of the exchange strength and even the sign of the exchange parameters, i.e. ferromagnetic or antiferromagnetic exchange interactions, depends much on details of the models \cite{Chaloupka02, Rau02, Katukuri, Yamaji, Sizyuk}. (This difficulty is also shared with $\alpha$-RuCl$_{3}$\cite{Banerjee, SHDo, Winter01}.) Inelastic neutron scattering (INS) is a traditional method to determine the exchange interactions by comparing measured dispersive magnetic excitations with the spin-wave theory. Only powder INS has been successfully performed for Na$_{2}$IrO$_{3}$ and $\alpha$-Li$_{2}$IrO$_{3}$ due to unavailability of large-volume single crystals and the high absorption of neutrons by Ir \cite{SKChoi01, SKChoi02}. An alternative way to probe magnon dispersion is resonant inelastic x-ray scattering (RIXS). In a RIXS investigation of a Na$_{2}$IrO$_{3}$ single crystal, a dispersive magnetic excitation was observed, but poor experimental resolution prevented the authors from determining the most relevant magnetic model Hamiltonian \cite{Gretarsson}. Another approach has been to employ quantum chemistry calculation or band calculations from first principles \cite{Yamaji, Foyevtsova, Nishimoto}. However, the sign and strength of the interactions depend on structural parameters sensitively, making it difficult to draw any firm conclusions. 

Here, we report a RIXS study that resolves momentum dependence of the spectrum associated with optical magnons in a single-crystal sample of $\alpha$-Li$_{2}$IrO$_{3}$, which exhibits a counter-rotating spiral magnetic order \cite{Williams}. Study of the dispersion relation of such a spiral magnet is helpful for elucidating the magnetic Hamiltonian, since the spiral magnetic ground state imposes strong additional constraints on the model parameters \cite{Williams, Biffin01, Biffin02, Kimchi, Stavropoulos, Baez}. Momentum- and temperature-dependences of the RIXS spectrum in the energy region 20-25 meV point to the presence of optical magnon modes. A comparison with the spin-wave theory suggests that this energy scale is largely determined by the ferromagnetic Kitaev interaction of order 20 meV, while the Heisenberg interaction is much smaller ($\sim$4 meV). Such a large Kitaev interaction also means that the short-ranged spiral magnetic correlation persists up to very high temperature (at least up to 5.5 times $\textit{T}_{N}$), which we confirm by diffuse magnetic scattering. We also observe that the $C_3$ symmetry is not recovered even at this high temperature, indicating that the anisotropy among the \textit{x}, \textit{y}, and \textit{z} bonds is not negligible. 

The $\alpha$-Li$_{2}$IrO$_{3}$ single crystals were grown by a vapor transport growth method \cite{Lin, Freund}. The resonant magnetic x-ray diffraction and the diffuse magnetic scattering data were collected at the 6ID-B,C beamline and the 27ID-B beamline of the Advanced Photon Source, respectively. The RIXS data were obtained at the ID20 beamline of the European Synchrotron Radiation Facility \cite{Sala01, Sala02}. Both diffuse magnetic scattering and RIXS instrumentations employ the Rowland circle configuration with the 2$\theta$ close to 90$\degree$ in the horizontal scattering geometry in order to detect the magnetic scattering by reducing Thompson scattering. A spherical diced Si(844) analyzer is used to achieve energy resolution of 24 meV (FWHM). The analyzer size (diameter: 6.35 cm) covers the solid angle (Rowland circle radius: 2 m) corresponding to momentum resolution of 0.1 r.l.u. along the $\textit{h}$ direction. The incident x-ray energy was tuned to 11.216 keV near the Ir $L_{3}$ edge for all of the resonant x-ray scattering.

We confirm the quality of the single crystal specimen first by the resonant magnetic x-ray diffraction. Figures 1(c)\&(d) show a magnetic Bragg peak at $\textit{\textbf{Q}}_{\textbf{mag}}$ = $\textit{\textbf{q}}_{\textbf{mag}}$ + (0, 0, 6) where $\textit{\textbf{q}}_{\textbf{mag}}$ = (0.32, 0, 0) scanned along the \textit{l} direction and the temperature dependence of the integrated intensity, respectively. These results are consistent with the counter-rotating spiral magnetic order \cite{Williams} below $\textit{T}_{N}$ = 12 K. The single crystal could in principle have two other twin domains that are rotated by $\pm$120$\degree$ about the \textit{l} direction. If these twins were present, they would give rise to magnetic Bragg reflections at (-0.165, 0.48, 6.16) and (-0.165, -0.48, 6.16), respectively. (These two domains do not have the same \textit{l} values in the monoclinic structure notation.) However, we find no intensity at these positions (Fig. 1(c)), indicating that the investigated sample region consists of a single magnetic and structural domain.

Figure 2(a) shows RIXS spectra at 4 K for different \textit{l} values of a magnetic Bragg spot, (\textit{h}, \textit{k}) = -$\textit{\textbf{q}}_{\textbf{mag}}$ + (-2, -2). The elastic magnetic signal dominates at the magnetic Bragg spot (where \textit{l} = 6) and its width confirms the energy resolution $\sim$24 meV (FWHM). Moving away from \textit{l} = 6 toward 6.42, inelastic spectral weight peaked at 25 meV becomes more prominent as the elastic signal decreases substantially. We note that this inelastic feature is independent of \textit{l} reflecting the two dimensional (2D) nature of the excitation that is confined in the honeycomb lattice layer. In addition, this inelastic spectrum remains almost unchanged for temperatures slightly above $\textit{T}_{N}$, e.g. at 19 K shown in Fig. 2(b), suggesting the presence of the 2D magnetic correlation above $\textit{T}_{N}$.

The inelastic spectrum changes substantially at very high temperatures. Figures 2(c)\&(d) compare the RIXS spectra at 4 K and 260 K at two selected \textbf{\textit{q}} values. Both spectra at 260 K show enhancement of spectral weight in the energy gain-side (energy transfer $<$ 0) possibly due to anti-Stokes contributions from either phonons \cite{Gretarsson, Vale} or magnetic excitations \cite{Revelli}. On the other hand, the energy-loss feature peaked around 20-25 meV is no longer prominent, and its spectral weight is reduced even after considering the Bose-Einstein thermal population factor. This observation suggests that the RIXS spectra at 4 K are dominated by magnetic excitations whose spectral weight diminishes at 260~K. We note that a recent RIXS study attributed this high-temperature background to a continuum of magnetic excitations attributed to Majorana fermions \cite{Revelli}. Here, we only focus on the momentum dependence of the magnetic excitations themselves in the magnetically ordered phase at 4~K, whose temperature dependence is distinct from that studied in Ref. 46. 

We examine the RIXS spectra at (\textit{h}, 0). Figure 3(a) presents the spectra at selected \textbf{\textit{q}} values, showing the momentum dependence of the spectral weight and profile. We first fit the energy-gain side with the instrumental resolution function and then subtract this elastic background (black line) at -$\textit{\textbf{q}}_{\textbf{mag}}$ (middle panel). The remaining inelastic spectrum (circles) shows a peak at 25 meV that is free from the gapless acoustic magnon. The raw spectra at \textbf{\textit{q}} = (0, 0) and (-0.66, 0) have no such substantial elastic background, but pronounced inelastic peaks have shifted to lower energy $\sim$20 meV (top and bottom panels), thus expressing a feature of momentum dependence. It is noted that these raw spectra exhibit an asymmetric energy profile that is hard to fit with symmetric Gaussian/Lorentizan profiles. We obtained the best fit by a combination of an elastic background and an inelastic spectrum represented by a damped harmonic oscillator model (Fig. 3(a)). Figures 3(b)\&(c) display pseudo-color scale intensity maps of the raw RIXS spectra and the background-subtracted spectra, respectively, as a function of momentum and energy transfers. The inelastic peak positions (triangles) obtained by fitting with the damped harmonic oscillator model is overlaid in Fig. 3(c). These positions are close to (at most $\sim$2 meV higher than) the inelastic peak energies of the raw spectra in Fig. 3(b), indicating that the momentum dependence is present whether the elastic contribution is considered or not. The momentum dependence of the RIXS spectra is symmetric for $\pm$\textit{h}, which is a characteristic of magnon bands and also confirmed in another Brillouin zone centered at (-2, -2), as shown in Fig. 3(d) (see also Section 3 in Supplemental Material \cite{SM}).

To gain quantitative understanding of the observed momentum-dependence, it is compared with a linear spin-wave theory calculation. To this end, we employed a generic nearest-neighbor spin model \cite{Rau02,Rau03};
\begin{align*}
H& = \sum_{\langle ij\rangle_{\gamma} } 
[K S_i^\gamma S_j^\gamma+\Gamma(S_i^\alpha S_j^\beta+S_i^\beta S_j^\alpha)\\
&+\Gamma'(S_i^\alpha S_j^\gamma+S_i^\gamma S_j^\alpha+S_i^\beta S_j^\gamma+S_i^\gamma S_j^\beta) + J {\mathbf S_i} \cdot {\mathbf S_j}],
\end{align*}
where $\langle ij\rangle_{\gamma}$ denotes a nearest-neighbor $\gamma$-bond ($\gamma \neq \alpha \neq \beta \in (x, y, z)$; See Fig. 1(a)), \textit{K} is the Kitaev interaction, $\Gamma$ is symmetric off-diagonal exchange interaction, $\Gamma$' is the additional exchange arising from trigonal distortion of the octahedron, and \textit{J} is the usual isotropic Heisenberg exchange interaction. For simplicity, we consider neither longer-range interactions nor bond-anisotropy. The counter-rotating spiral order is stabilized principally by a ferromagnetic \textit{K} and antiferromagnetic $\Gamma$, while both $\Gamma$' and \textit{J} play minor roles \cite{Rau02, Rau03}. The possible exchange parameter ranges have been explored with the constraint that stabilizes the counter-rotating spiral ground state while $\textit{\textbf{q}}_{\textbf{mag}}$ is approximated as (1/3, 0, 0) \cite{Stavropoulos}. Within this constraint, the condition $|\textit{K}| > |\Gamma|$ reproduces the observed dispersion of magnetic excitations \cite{SM}. (The spin wave dispersion changes slightly depending on the ratio of $\Gamma$' and \textit{J}.)

The magnon dispersion relation and its intensity plotted in Fig. 3(f) were simulated with a parameter set: \textit{K} = -25 meV, $\Gamma$ = 17.4 meV, $\Gamma$' = 3.2 meV, and \textit{J} = 4.4 meV. Figure 3(e) shows the calculated intensity plot convoluted with instrumental resolution function (Gaussian functions with the momentum width $\Delta$\textit{\textbf{q}} $\sim$0.1 r.l.u. and the energy width $\Delta$E $\sim$24 meV) (see Fig. S4 in Supplemental Material \cite{SM}). The gapless acoustic magnon mode centered around $\textit{\textbf{q}}_{\textbf{mag}}$ dominates the spectrum, but its intensity is drastically diminished moving away from $\textit{\textbf{q}}_{\textbf{mag}}$. Because of this strong momentum dependence and the finite resolution, we could not distinguish this mode from the elastic magnetic intensity. We also notice optical magnon-like dispersion in the calculation that is consistent with our observation shown in Fig. 3(c). According to the magnon dispersion presented in Fig. 3(f), this dispersive feature is explained by a superposition of several optical magnon modes that have momentum-dependent scattering intensities. However, the coarse energy resolution used here does not allow us to make a sharp statement about the interaction parameters, and different parameter sets can account for our results equally well. For example, the dispersion can be explained with other spin models that take into account the difference in the exchange interactions among \textit{x}, \textit{y}, and \textit{z} bonds \cite{Kimchi, Winter02}. Although all the exchange interaction parameters may vary upon specific models, we find that dominant Kitaev interaction of order 20 meV is required to describe our RIXS results.

Next, we discuss diffuse magnetic scattering. Figure 4(a) depicts a diffuse magnetic x-ray scattering map projected on the \textit{hk} plane at 20 K. (This 2D magnetic correlation is independent of the \textit{l} value.) It shows a pair of diffuse features at nearly the same \textit{h} and \textit{k} as $\textit{\textbf{Q}}_{\textbf{mag}}$, which suggests that the spiral magnetic correlation persists well above $\textit{T}_{N}$ despite a short correlation length. The correlation length is estimated to be $\sim$3 unit cells (1/HWHM, $\sim$16 \AA{}) of the honeycomb lattice and shows little temperature dependence between 16~K and 66~K (Fig. 4(c)). We tracked the diffuse scattering intensity as a function of temperature, monitoring the peak intensities at the elastic line of the RIXS spectra at $\textit{\textbf{Q}}_{\textbf{mag}}$ (Fig. 4(d)). The elastic intensity in the temperature range between 77 K and 250 K is temperature-independent, and thus regarded as an elastic background. Note that the spectrum at 14 K has the most pronounced spectral weight at the elastic line, which gradually decreases and converges to the one at 77 K. This temperature-dependent spectral weight is regarded as the diffuse magnetic intensity and is present up to at least $\sim$5.5$\textit{T}_{N}$, which is close to 100 K where the magnetic susceptibility deviates from the Curie-Weiss behavior \cite{Singh02}. Given the dominant Kitaev interaction, this observation corroborates the significant magnetic fluctuation in $\alpha$-Li$_{2}$IrO$_{3}$ and reflects a possible link to the Kitaev QSL state.

It is interesting to compare the diffuse scattering result with that of the sister compound Na$_{2}$IrO$_{3}$. This compound exhibits three pairs of diffuse intensities above $\textit{T}_{N}$, each corresponding to one of three short-range zigzag magnetic orders that propagate along the directions rotated by $\pm$120$\degree$ from one another, i.e. \textit{x}, \textit{y}, and \textit{z} bonds \cite{Chun}. These three short-range orders are equally populated once the long-range order melts above $\textit{T}_{N}$, since all the exchange interactions associated with each bond have similar strength to those of the other bonds.

$\alpha$-Li$_{2}$IrO$_{3}$ does not exhibit such three pairs associated with the \textit{C$_{3}$} symmetry. Figure 4(e) shows the RIXS spectra at $\textit{\textbf{Q}}_{\textbf{mag}}$ and $\pm$120$\degree$ rotated positions at 20 K and an azimuthal angle $\Psi$ = 150$\degree$. It is evident that only $\textit{\textbf{Q}}_{\textbf{mag}}$ presents substantial elastic spectral weight corresponding to the diffuse magnetic scattering. Figure 4(f) shows that this spectral weight closely follows the $\Psi$-dependence of summed $\pi$-$\pi$' and $\pi$-$\sigma$' channels simulated for the long-range order \cite{Williams}, indicating that the short-range order preserves the magnetic anisotropy. If the other two pairs were to coexist, they would exhibit a $\Psi$-dependence shifted by 120$\degree$ and 240$\degree$, and in turn, at least one of them (e.g. the cyan line in Fig. 4(e)) is expected to display considerable elastic spectral weight at $\Psi$ = 150$\degree$, which is inconsistent with our observation. The absence of the other two short-range correlations points to sizeable departure from the ideal \textit{C$_{3}$} symmetry \cite{Winter02}.

In summary, we report a resonant inelastic x-ray scattering study of magnetic excitations in a single crystal of the honeycomb lattice iridate $\alpha$-Li$_{2}$IrO$_{3}$. The energy of the spectrum in the range 20-25 meV is highest at the magnetic Bragg peak positions. This momentum dependence is attributed to optical magnon modes and a spin-wave analysis reveals a dominant ferromagnetic Kitaev interaction of order 20 meV. Our diffuse magnetic scattering study above $\textit{T}_{N}$ shows a persistent magnetic correlation up to $\sim$5.5$\textit{T}_{N}$, and in turn suggests significant magnetic fluctuation associated with the Kitaev interactions. The diffuse scattering pattern deviates from the \textit{C$_{3}$} symmetric behavior, pointing to a degree of bond anisotropy in $\alpha$-Li$_{2}$IrO$_{3}$, not found in Na$_{2}$IrO$_{3}$.

Work in the Materials Science Division at Argonne National Laboratory (crystal growth and resonant magnetic x-ray diffraction) was supported by the US Department of Energy, Office of Science, Basic Energy Sciences, Materials Science and Engineering Division. Work at the University of Toronto (RIXS, data analysis, and spin wave analysis) was supported by the Natural Science and Engineering Research Council (NSERC) of Canada. P. P. S. and H.-Y. K. were supported by the NSERC of Canada Discovery Grant No. 06089-2016, and H.-Y. K. acknowledges funding from the Canada Research Chairs Program. Use of the Advanced Photon Source at Argonne National Laboratory was supported by the US Department of Energy, Office of Science, under Contract No. DE-AC02-06CH11357. Work at Pohang Accelerator Laboratory (data analysis and discussion) was supported by National Research Foundation of Korea (2019R1C1C1010034 and 2019K1A3A7A09033399). B. J. K. acknowledges support by IBS-R014-A2. 

Note added.-Recently, we noted that ferromagnetic characteristics of the Kitaev exchange interaction have been identified in $\alpha$-RuCl$_{3}$ using resonant magnetic x-ray diffraction (Ref. \cite{Sears}) and dispersive magnetic excitations in Na$_{2}$IrO$_{3}$ is studied using high resolution RIXS (Ref. \cite{JHKim}). The energy scale of Kitaev interaction for both compounds is contingent on details of the specific magnetic Hamiltonian model applied in the analysis. 


\begin{figure}[p] 
	\centering
	\includegraphics[width=5in]{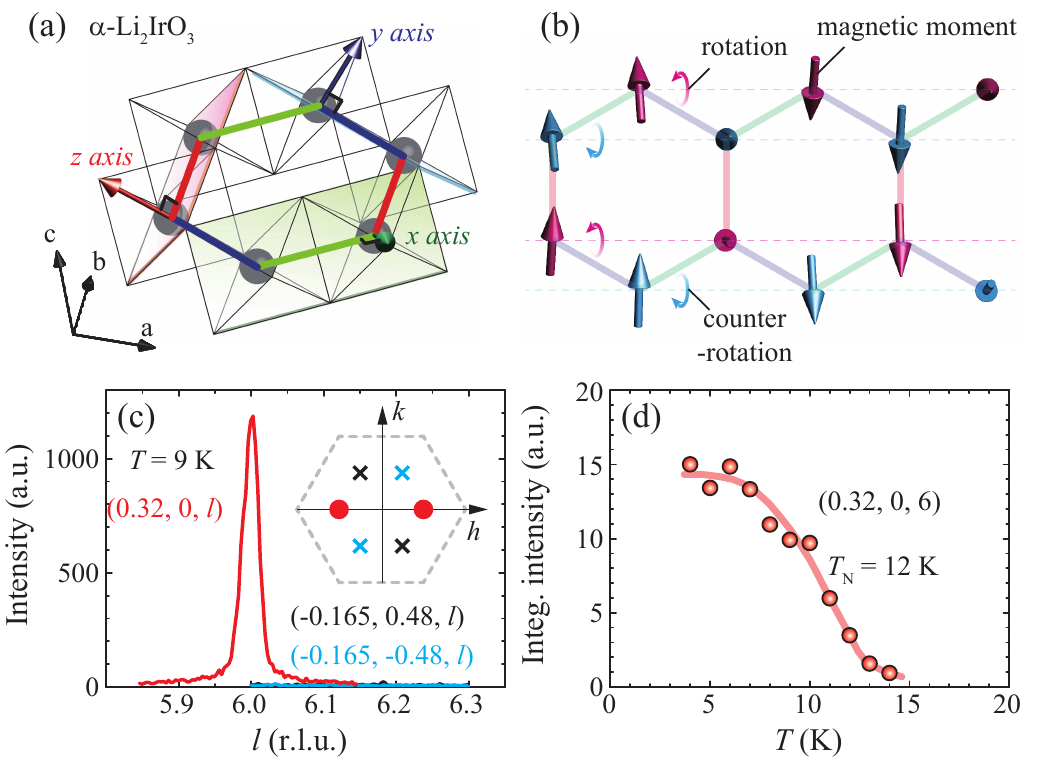}
	\caption{
		(a) Honeycomb lattice of $\alpha$-Li$_{2}$IrO$_{3}$ formed by Ir$^{4+}$ ions (spheres) inside the edge-shared oxygen octahedra. The other ions are not shown for clarity. The \textit{x}, \textit{y}, and \textit{z} bonds denoted with green, blue, and red lines are normal to the local \textit{x}, \textit{y}, and \textit{z} axes of the octahedron, respectively. (b) Counter-rotating spiral magnetic order. The magnetic moments on each dashed line ($\parallel$ \textit{a} axis) rotate about the line. The rotation directions are opposite between the adjacent lines. (c) The \textit{l}-scans at the magnetic Bragg peak positions for three possible magnetic domains. Each pair of the same color-coded circles (red) and crosses (black and cyan) inside the hexagonal Brillouin zone depict the positions for the different domains (inset). (d) Temperature dependence of the integrated intensity at $\textit{\textbf{Q}}_{\textbf{mag}}$ stemming from the magnetic order below $\textit{T}_{N}$ = 12 K.
	}
	\label{fig:example}
\end{figure}

\begin{figure}[p] 
	\centering
	\includegraphics[width=5in]{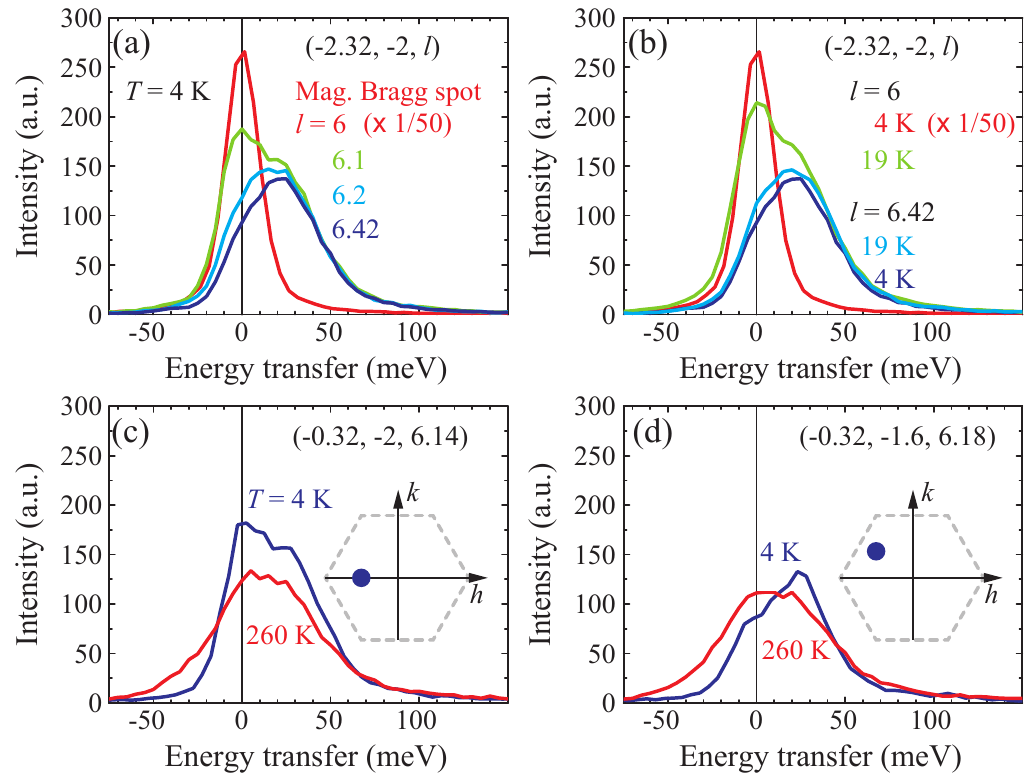}
	\caption{
		(a) The \textit{l}-dependence of RIXS spectra at -$\textit{\textbf{q}}_{\textbf{mag}}$ + (-2, -2) at 4 K. The spectrum of the magnetic Bragg spot is reduced by 50 for comparison with those of the other \textit{l} values. (b) The RIXS spectra of \textit{l} = 6 and 6.42 below and slightly above $\textit{T}_{N}$, i.e. 4 K and 19 K, respectively. (c, d) The RIXS spectra at (\textit{h}, \textit{k}) = (-0.32, -2) and (-0.32, -1.6) marked with the circles inside another Brillouin zone (inset) below and exceedingly above $\textit{T}_{N}$, i.e. 4 K and 260 K.
	}
	\label{fig:example}
\end{figure}

\begin{figure}[p] 
	\centering
	\includegraphics[width=5in]{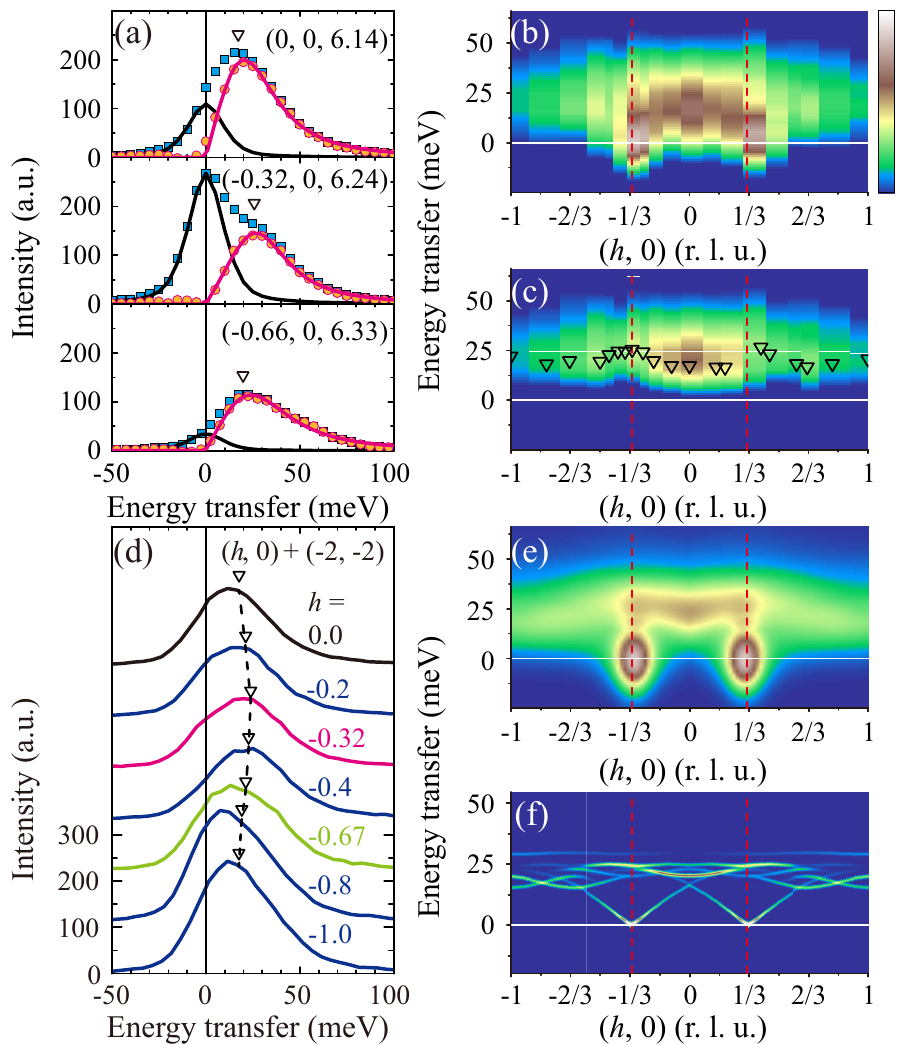}
	\caption{
		(a) RIXS spectra at \textit{\textbf{q}} = (0, 0) (top), -$\textit{\textbf{q}}_{\textbf{mag}}$ (middle), and (-0.66, 0) (bottom). The raw spectrum and elastic background are shown with the squares and black line, respectively. The circle and pink line depict the spectrum subtracted by the elastic background and fitting result from a damped harmonic oscillator (DHO) model, respectively. Intensity maps of the raw RIXS spectra (b) and the spectra subtracted by the elastic background (c) at 4 K along (\textit{h}, 0). Red dashed lines correspond to $\pm$$\textit{\textbf{q}}_{\textbf{mag}}$. Open triangles depict the peak centers obtained by fitting with the DHO model. (d) RIXS spectra along (\textit{h}, 0) in another Brillouin zone centered at (-2, -2). The \textit{l} value ranges from 6.42 to 6.36. The dotted line is a guide to the eye. (e) Intensity map of the calculated magnon dispersion after convolution with the energy and momentum instrumental functions. (f) Magnon dispersion relations along (\textit{h}, 0) simulated by the linear spin wave theory based on the \textit{K}-$\Gamma$-$\Gamma$'-\textit{J} model.
	}
	\label{fig:example}
\end{figure}

\begin{figure}[p] 
	\centering
	\includegraphics[width=5in]{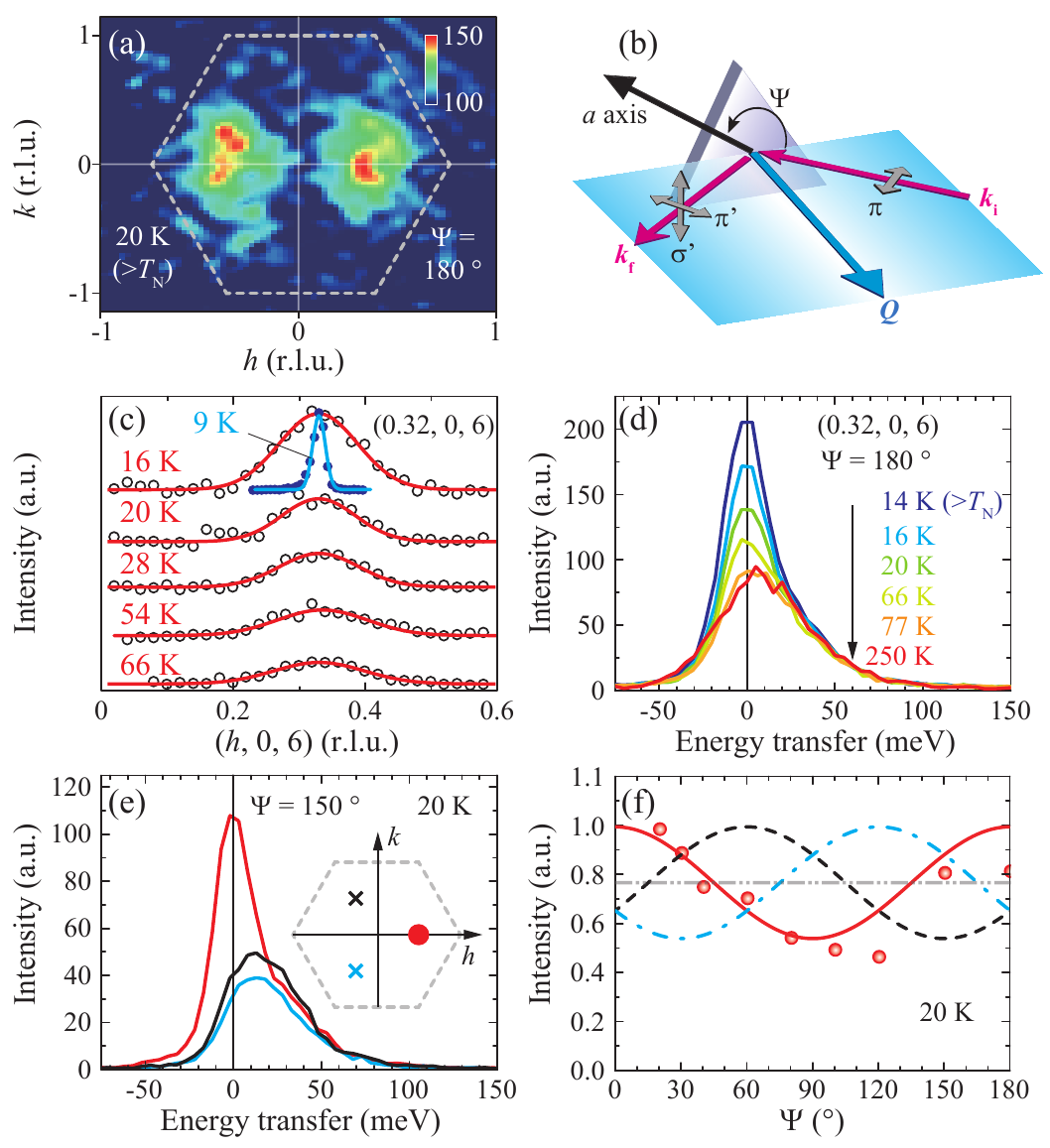}
	\caption{
		(a) Intensity map of diffuse magnetic scattering on the \textit{hk} plane at 20 K. (b) Illustration of the horizontal scattering geometry. The scattering plane is defined by the incident ($\textit{\textbf{k}}_{\textbf{i}}$) and outgoing ($\textit{\textbf{k}}_{\textbf{f}}$) wavevectors. The incident ($\pi$) and outgoing x-ray polarizations ($\pi$', $\sigma$') are shown with green arrows. The azimuthal $\Psi$ is defined as the angle between the \textit{a} axis and the scattering plane (the curved black arrow). (c) Temperature(\textit{T})-dependence of the diffuse scattering along (\textit{h}, 0, 6). The data are vertically shifted for clarity. The magnetic Bragg peak (blue) is scaled down for comparison. (d) \textit{T}-dependence of the RIXS spectra at $\textit{\textbf{Q}}_{\textbf{mag}}$. (e) RIXS spectra at $\textit{\textbf{Q}}_{\textbf{mag}}$ (red line) and $\pm$120$\degree $ rotated positions about the \textit{l} direction (black and cyan lines) at 20 K and $\Psi$ = 150$\degree$. The same color-coded spots are denoted inside the Brillouin zone (inset). (f) Elastic spectral weight at $\textit{\textbf{Q}}_{\textbf{mag}}$ for selected $\Psi$ (circles). Solid and dashed curves display the simulated $\Psi$-dependence of summed $\pi$-$\pi$' and $\pi$-$\sigma$' channels for the counter-rotating spiral orders at the same color-coded positions in the inset of (e). The average of these three curves is denoted as grey line. 
	}
	\label{fig:example}
\end{figure}
\end{document}